\documentclass[preprintnumbers,amsmath,amssymb,floatfix,11pt,prd,onecolumn,showpacs,superscriptaddress,nofootinbib]{revtex4}
\usepackage{graphicx}
\usepackage{epsfig}
\usepackage{bm}
\usepackage{amsfonts}

\begin{document}

\title{Resolution of dark matter problem in f(T ) gravity}

\author{\textbf{ Mubasher Jamil}} \email{mjamil@camp.nust.edu.pk}
\affiliation{Center for Advanced Mathematics and Physics (CAMP),\\
National University of Sciences and Technology (NUST), H-12,
Islamabad, Pakistan}
\affiliation{Eurasian International Center
for Theoretical Physics, Eurasian National University, Astana
010008, Kazakhstan}
\author{\textbf{ D. Momeni}}
\email{d.momeni@yahoo.com}
 \affiliation{Eurasian International Center
for Theoretical Physics, Eurasian National University, Astana
010008, Kazakhstan}
\author{\textbf{ Ratbay Myrzakulov}}
\email{rmyrzakulov@gmail.com}\affiliation{Eurasian International Center
for Theoretical Physics, Eurasian National University, Astana
010008, Kazakhstan}

\begin{abstract}
\textbf{Abstract:} In this paper, we attempt to resolve the dark matter
problem in f(T) gravity. Specifically, from our model
we successfully obtain the flat rotation curves of galaxies
containing dark matter. Further, we obtain the density profile
of dark matter in galaxies. Comparison of our analytical
results shows that our torsion-based toy model for dark matter
is in good agreement with empirical data-based models.
It shows that we can address the dark matter as an effect of
torsion of the space.

\end{abstract}
 \maketitle

\newpage

%%%%%%%%%%%%%%%%%%%%%%%%%%%%
\section{introduction}
%%%%%%%%%%%%%%%%%%%%%%%%%%%%
As we know Dark (non-luminous and non-absorbing) Matter (DM) is an
old idea even before the dark energy problem,  which is causing the
accelerating expansion of the Universe in the large scale
\cite{dm_bertone,dm_hist}. The most accepted observational evidence
for existence of such component comes from the astrophysical
measurements of several  galactic rotation curves. From the view of
classical mechanics, we expect that the rotational velocity
$v_\varphi$ of any astrophysical  object moving in  a stable (quasi
stable) Newtonian circular orbit with radius $r$ must be in the form
$$
v_{\varphi}(r) \propto \sqrt{
  M(r) / r},
  $$
where $M(r)$ is identified as the mass (effective mass) profile
thoroughly inside the orbit. For many spiral and elliptical galaxies
this velocity $v_{\varphi}$ remains  approximately constant for
large galactic radii, for instance in the Milky way galaxy, $v
\simeq 240 km/s$. This estimation is valid only near the position of
our solar system. There is a lower bound on the DM mass density,
$\Omega_{\rm DM} \approx 0.1$ from phenomenological particle
physics. There are different kinds of the dark matter \cite{DM}. To
solve the DM problem several proposals were introduced. The problem
can be interpreted as an effect of the extra dimensions in a
cosmological special relativity (CSR) model, proposed by Carmelli
\cite{moshe}. From particle physics view as LSP in supersymmetric
theories or LKP in higher dimensional theories in which the SM
(standard model) predicts some extra dimensions. Stability condition
of any candidate for dark matter is very important problem which
must be checked. For example for stabilization checking in SUSY
(super symmetry) we must check the validity of R- parity and in
supergravities alternatives we must following the KK parity. In
brief, ``\emph{Any candidate for dark matter need not be stable if
its abundance at the time of its decay is sufficiently small}".
There are several classical candidates for dark matter, as perfect
fluid models \cite{models} and as the geometrical modifications of
the Einstein-Hilbert action, for example $R^2$ modification of the
usual Einstein gravity \cite{r2} or in the anisotropic and
dfiffeomorphism invariance model of Horava-Lifshitz as an
integration constant \cite{hl}.

In this Letter we focus on the mechanism of the $f(T)$ gravity and
show that in the context of this new proposed non-Riemannian
extension of the general relativity (GR), it is possible to explain
the rotation curves of the galaxies without introducing dark matter.
Our plan in this letter is: In section II we propose the basis of
the $f(T)$ gravity. In section III, we investigate the spherically
symmetric solutions of the model. In section IV we solve the
equations and show that the rotation curve of the galaxies in this
toy model of the spherically-symmetric-static model can be recovered
by the effects of the torsion alone. In section V we obtain the halo
density profile and compare it with two well known astrophysical
models. We conclude in the final section.

%%%%%%%%%%%%%%%%%%%%%%%%%%%%%%%%%%%%%%%%
\section{Formalism of $f(T)$ gravity}
%%%%%%%%%%%%%%%%%%%%%%%%%%%%%%%%%%%%%%%%

A gauge theory of gravity is based on the  equivalence principle.
For example, $SL(2,C)$ gauge theory on the gravitational field can
be used for quantization of this fundamental force \cite{carmeli2}.
We are working with a curved manifold for the construction of a
gauge theory for gravitational field. It is not necessary to use
only the Riemannian manifolds. The general form of a gauge theory
for gravity, with metric, non-metricity and torsion can be
constructed easily \cite{smalley}. If we relax the non-metricity,
our theory is defined on Weitzenb\"{o}ck spacetime, with torsion and
with zero local Riemann tensor $R_{\alpha\beta\gamma}^{\delta}=0$.
In this theory, which is called teleparallel gravity, we use a
non-Riemannian spacetime manifold. The dynamics of the metric
determined using the torsion $T$. The basic quantities in
teleparallel or the natural extension of it, namely $f(T)$ gravity
is the vierbein (tetrad) basis  $e^{i}_{\mu}$
\cite{ff,linder,darabi}. This basis is an orthonormal, coordinate
free basis, defined by the following equation
\begin{eqnarray}\nonumber
g_{\mu\nu}=e_{\mu}^{i}e_{\nu}^j \eta_{ij}\,.
\end{eqnarray}
This tetrad basis must be orthonormal and $\eta_{ij}$ is the
Minkowski flat tensor. It means that $e^{i}_{\mu}e^{\mu
}_j=\delta^i_{j}$.
%$e^{i}_{\mu}e_{\mu}_{(j)}=\delta^i_{j}$.
One suitable form of the action for $f(T)$ gravity in Weitzenb\"{o}ck
spacetime is given by \cite{darabi}
\begin {equation}\label{a-1}
S=\int d^{4}xe\Big(\frac{1}{16\pi}[T+f(T)]+L_{m}\Big)\,,
\end{equation}
% Here
where  $f$ is an arbitrary function, $e=\det(e^{i}_{\mu})$. Here $T$ is defined by
\begin{equation}\nonumber
T=S^{\:\:\:\mu \nu}_{\rho} T_{\:\:\:\mu \nu}^{\rho}\,,
\end{equation}
with
$$
T_{\:\:\:\mu \nu}^{\rho}=e_i^{\rho}(\partial_{\mu}
e^i_{\nu}-\partial_{\nu} e^i_{\mu})\,,
$$
$$
S^{\:\:\:\mu \nu}_{\rho}=\frac{1}{2}(K^{\mu
\nu}_{\:\:\:\:\:\rho}+\delta^{\mu}_{\rho} T^{\theta
\nu}_{\:\:\:\theta}-\delta^{\nu}_{\rho} T^{\theta
\mu}_{\:\:\:\theta})\,,
$$
where the asymmetric tensor (which is also called the contorsion tensor) $K^{\mu \nu}_{\:\:\:\:\:\rho}$ reads
$$
K^{\mu \nu}_{\:\:\:\:\:\rho}=-\frac{1}{2}(T^{\mu
\nu}_{\:\:\:\:\:\rho}-T^{\nu \mu}_{\:\:\:\:\:\rho}-T^{\:\:\:\mu
\nu}_{\rho})\,.
$$
The equation of motion derived from the action, by varying the action with respect to the $e^{i}_{\mu}$,
is given by
\begin{eqnarray}
e^{-1}\partial_{\mu}(e S^{\:\:\:\mu
\nu}_{i})(1+f_T)-e_i^{\:\lambda}T_{\:\:\:\mu
\lambda}^{\rho}S^{\:\:\:\nu \mu}_{\rho}f_T +S^{\:\:\:\mu
\nu}_{i}\partial_{\mu}(T)f_{TT}-\frac{1}{4}e_{\:i}^{\nu}
(1+f(T))=4 \pi  e_i^{\:\rho}T_{\rho}^{\:\:\nu}\label{eom}
\end{eqnarray}
where
$T_{\mu\nu}=e^{a}_{\mu}T_{a\nu}$ is the energy-momentum tensor for matter sector of the Lagrangian $ L_m$, it is defined using
$$
T_{a\nu}=\frac{1}{e}\frac{\delta L_m}{\delta e^{a\nu}}
$$
The covariant derivatives compatible with the metricity
$g^{\mu\nu}_{;\mu}=0$. It is a straightforward calculation to show
that  (\ref{eom}) is reduced to the Einstein gravity when $f(T)=0$.
This is the equivalence between the teleparallel theory and the
Einstein gravity \cite{T}. Note that teleparallel gravity is not
unique, since it can either be described by any Lagrangian which
remains invariance under the local or global Lorentz $SO(3, 1)$
group \cite{tegr}.

We mention here that a general Poincare gauge invariance model for
gravity (is so called Einstein-Cartan-Sciama-Kibble (ECSK))
previously reported in the literatures \cite{ECSK}. Specially, in
the framework of the Poincaré gauge invariant form of the
ECSK,theory the \textit{notions of ``dark matter'' and ``dark
energy'' play the role similar to that of ``ether'' in physics
before the creation of special relativity theory by Einstein}
\cite{plb}.  In this Letter we focus only on $f(T)$ models, without
curvature and with non zero torsion. \textbf{We would remark that an
attempt to explain the flat rotation curve of galaxies has been made
earlier in the framework of ECSK theory \cite{ml}.}

%%%%%%%%%%%%%%%%%%%%%%%%%%%%%%%%%%%%%%%%%%%
\section{ Spherically Symmetric geometry}
%%%%%%%%%%%%%%%%%%%%%%%%%%%%%%%%%%%%%%%%%%%%%
As is well-known that a typical spiral galaxy contains two
forms of matter: luminous matter in the form of stars and stellar
clusters which are found in the galactic disk while another is dark
matter which is generally found in the galactic halo and
encapsulates the galaxy disk. In the early Universe, the DM played
crucial role in the formation of galaxies when the dense
concentration of DM lied in the galactic centers which helped in the
accumulation of more dust and gas to form proto-galaxies. In the
later stages of galactic evolution, the DM slowly drifted towards
the outer regions of the galaxies forming a huge (but less
concentrated/dense) DM halos. Although the precise form of
distribution of dark matter in the halos is not known, we assume
that the spatial geometry of galactic halo is spherically symmetric.
Moreover, the dark matter halo is isotropic: the spherical DM halo
expands (hypothetically) only radially while having no tangential or
orthogonal motions relative to the radial one. From the point of
view of Grand Unified Theories (GUT), the most likely candidate of
DM is neutralino which is a weakly interacting massive particle
\cite{claus} with additional minor contribution form primordial
black holes $\Omega_{\text{PBH}}=10^{-8}$ which were formed in the
early Universe and are also candidate for other violent cosmic
events like Gamma Ray Bursts \cite{nayak}. Note that we are not
interested in any particular form of DM and deal only with its
characteristic role in the rotation of galactic disks. Our
theoretical model suggests that the flat galactic rotation curves
can be explained in terms of torsion of space without invoking dark
matter. In other words, the huge DM halo is nothing but mysterious
and elusive torsion of space. Now we construct a model for galaxy, based on the above assumptions.
The metric of a static spherically symmetric (SSS) spacetimes can be described, without loss of generality, as

\begin{equation}
ds^{2}=e^{a(r)}dt^{2}-e^{b(r)}dr^{2}-r^{2}\left(d\theta^{2}+\sin^{2}\theta d\phi^{2}\right)\label{g}\; .
\end{equation}
This form of metric is updated by the Schwarzschild gauge, and is useful to construction of a toy model for galaxy. In order to re-write the metric  (\ref{g}) into the invariant form under the Lorentz transformations, we use the tetrad matrix \cite{h}
\begin{eqnarray}
\left\{e^{i}_{\;\;\mu}\right\}= diag \left\{e^{a(r)/2},e^{b(r)/2},r,r\sin\theta\right\}\label{tetra}\; .
\end{eqnarray}
Although the $f(T)$ gravity is not local Lorentz invariance
\cite{prl}, but we can impose local invariance symmetry on the
metric components. Here we note some remarks about the relation
between local Lorentz invariance of the $f(T)$ as a scalar
gravitational theory and choice of the tetrads, specially in the
case of spherically symmetric metric, given by (\ref{tetra}). As we
know, teleparallel gravity with $f(T)=T$ is local lorentz
invariance, for any set of the constants $c_{1,2}$, since it is
equivalence to the Einstein theory, it means the total Lagrangian of
the linear $T$ theory is equivalence to the Einstein plus a surface
boundary term which can be canceled in the derivation of the
equations of motion \cite{T}. Boundary terms like
$n^{\nu}T^{\mu}_{\mu\nu}(\text{T is torsion})$ have thermodynamical
meaning but are free of the dynamics. Torsion based $f(T)$ theory,
constructed from the tetrad basis $e^i_{\mu}$ must be local
invariance under a proper Lorentz transformations. But in this form
and with usual tetrad basis, it has been shown that, this invariance
breaks \cite{prl}. Recently (without any direct proof), the authors
of \cite{bohmer}, shown that with another choice of the tetrads
(they called ``good tetrad"), which  leads to the non diagonal
metric components, the restriction on the form of the $f(T)$ as a
linear $T$ theory, came from the equation
$$
f_{TT}=0,
$$
relaxes.  This relaxation can be interpreted as a rotation of the
tetrads basis. In this new tetrads basis, there are three free Euler
angels (all local functions of the coordinates $r,t$), and the
expression of the scalar torsion has some extra terms. In this case
the system of equations even for vacuum case is very complicated.
Further, they found, it is possible to recover the Schwarzschild-de
Sitter as an exact black hole solution in this theory, and test the
parameters of the model, using the usual tests as PPN and etc. There
are some points that must be clarified by those authors
\cite{bohmer} to support their conclusions, for example does this
local invariance help to the power counting renormalizability of the
$f(T)$ model as an alternative theory? There are many kinds of such
rotational transformations. How we preferred one kind of the
rotations from others ?  We will be back to these problems in a
forthcoming paper on this topic \cite{ren}. Any way in this letter
we want to explore the effect of the torsion for construction of a
geometrical model for DM. For this purpose, the usual tetrads are
enough.

 Using  (\ref{tetra}), one can  obtain  $e=\det{\left[e^{i}_{\;\;\mu}\right]}=e^{(a+b)/2}r^2 \sin\left(\theta\right)$, and  the  torsion scalar  in terms of  $r$ is given by
\begin{eqnarray}
T(r) &=& \frac{2e^{-b}}{r}\left(a^{\prime}+\frac{1}{r}\right)\label{te}\;
\end{eqnarray}
where the prime  ($^{\prime}$) denote the  derivative with respect
to  the radial coordinate $r$. The equations of motion  for an
anisotropic fluid are \cite{h}
\begin{eqnarray}
4\pi\rho &=& \frac{f}{4}-\left( T-\frac{1}{r^2}-\frac{e^{-b}}{r}(a'+b')\right)\frac{f_T}{2}\,,\label{dens} \\
4\pi p_{r} &=& \left(T-\frac{1}{r^2}\right)\frac{f_T}{2}-\frac{f}{4}\label{presr}\;, \\
4\pi p_{t} &=& \frac{f_T}{2}\left[\frac{T}{2}+e^{-b}\left(\frac{a''}{2}+\left(\frac{a'}{4}+\frac{1}{2r}\right) (a'-b')\right)\right]-\frac{f}{4}\label{prest},\\
& &\frac{\cot\theta}{2r^2}T^{\prime}f_{TT}=0\;\label{impos},
\end{eqnarray}
where $p_{r}$ and $p_{t}$  are the radial and tangential pressures respectively, $\rho$ is density profile. This last quantity is very importnant in our astrophysical predictions. . Here if we use from the "`good"' tetrads \cite{bohmer}, the out coming system becomes very complicated as the following:
\begin{eqnarray}
  4\pi\rho = \frac{f}{4} -\frac{f_T\,e^{-b}}{4r^2}( 2-2\,e^b+r^2e^b T-2r\,b' )
  -\frac{f_{TT}\,T'e^{-b}}{r} (1+e^{b/2}\sin\gamma)
\\
  4\pi p = -\frac{f}{4} +\frac{f_T\,e^{-b}}{4r^2}( 2-2\,e^b+r^2e^b T-2r\,a' t)\\
  f_{TT}\,T'\cos\gamma = 0 \label{ftt}\\
   f_T\,\dot b = 0\\
  \dot{b} [ e^b r^2 f_T + 2\,f_{TT}(1+e^{b/2}\sin\gamma)(2-2\,e^b+r^2e^b T+ 2r\, a') ]
  - 4r\,f_{TT}\,\dot{a}'( 1+e^{b/2}\sin\gamma)^2 =0
\\
  f_{TT}\Big[ -4\,e^a r\, T' -\dot{b}^2( 2-2\,e^b+r^2e^b T ) -2r\,a' (e^ar\,T'+\dot b^2)
  +4r \dot b\dot a' (1+e^{b/2}\sin\gamma) \Big]
  \\ \nonumber+f_T \Big[ 4\,e^a-4e^ae^b-e^ar^2a'^2+ 2\,e^ar\,b'
  +e^ar\,a'( 2+r\,b') -2r^2e^a a''
  - e^br^2 \dot a\dot b+e^br^2 \dot b^2 +2e^br^2\ddot b \Big] =0
\end{eqnarray}
Here $f_{TT}$ is the second derivatives of $f(T)$ and overdots
denote  differentiation with respect to $t$, the Euler angles is
$\gamma(r,t)$. We do not give any further discussion on this system,
we will follow the simple, tetrads basis, given by (\ref{tetra}).
Note that even in this new tetrads basis, the teleparallel case is
preserved by equation (\ref{ftt}).   Thus even by this strange basis
of tetrads, we can include the teleparallel case with linear $T$
behavior. There is a vast family of exact solutions for this system,
which has been investigated previously \cite{h}. We focus only on
the following possible solutions, which arise  from both equations
(\ref{impos},\ref{ftt}) :
\begin{eqnarray}
T^{\prime}&=&0\Rightarrow T=T_0\;,\\
f_{TT}&=&0\Rightarrow f(T)=c_2+c_1T\;,\\
T^{\prime}&=&0,f_{TT}=0\Rightarrow T=T_0,f(T)=f(T_{0});,
\end{eqnarray}
which always relapses into the particular case of teleparallel
Theory, with $f(T)$ a  constant or a linear function. We adopt this
linear teleparallel choice for our physical discussions about the
possible explanation of the DM in the context of the torsion based
gravity, $f(T)$.  In the next section, we will solve the above
equations (\ref{dens}), (\ref{presr}) and (\ref{prest}) for the
metric function $a(r)$. In the language of the $3+1$ decomposition
of the metrics, determining $a(r)$  is equivalent to finding the
lapse (or redshift) function $N(r)=e^{\frac{a(r)}{2}}$.

%%%%%%%%%%%%%%%%%%%%%%%%%%%%%%%%%%%%%%%%%%%%%%%%%
\section{Dark matter problem in $f(T)$ Gravity }
%%%%%%%%%%%%%%%%%%%%%%%%%%%%%%%%%%%%%%%%%%%%%%%
The quasi global solution for (\ref{g}) with  assumption
$a(r)=-b(r)$ and by imposing the isotropicity in the pressure
components $p_r=p_t$ is the Schwarzschild-(A)dS presented by
\cite{h}
$$
e^{a(r)}=e^{-b(r)}=1-\frac{c_0}{r}+\frac{c_1}{3}r^2.
$$
Obviously such trial metric can not be successful to  generating the
rotation curve of the spiral galaxies. Indeed this classical
solution leads the zero torsion $T=0$. From physical intuition we
know the DM problem must be comes from a non zero torsion, and
specially from a variable one, $T=T(r)$. Now we introduce an ansatz
for solution and choose
\begin{equation}
b(r)=c\label{b},
\end{equation}
where $c$ is an arbitrary  constant\footnote{Here $c$ does not
represent the speed of light}. Another choice is the polynomial form
for the lapse function $e^a$ \cite{farook}, but which such choices ,
independent from the origion which they come, the rotation curve
fixed by a desirable linear form, and it seems that such choices are
ad hoc and not physically acceptable. Further we choose another
ansatz  $c_1=1$ and $c_2=0$. It means $f(T)=T$. The model and the
field equations still remain scale invariant. The main reason for
choice of the metric function $b$ as a constant goes back to the
scale invariance of the system and further, we are interested to a
lapse function $N=e^a$ which can explain the flat rotation curve. We
discuss more on why we choose such a restricted gauge. Let us
consider the following static form of the metric
\begin{equation}
g_{\mu\nu}dx^\mu dx^\nu=h_{AB}dx^Adx^B-\Phi^2d\Sigma_2^2\label{g2}
\end{equation}
instead of (\ref{g}), which is a four dimensional dual of the
following renormalizable effective action, defined on the two
dimensional induced metric $h_{AB}(x,t),\ \ A,B=0,1$,
\begin{equation}
S=- \frac{1}{16\pi}\int d^2x \sqrt{-h}\Big[f(\Phi)T+\epsilon\Phi_{;A}\Phi^{;A}-U(\Phi)\Big],
\ \ \epsilon=\text{constant}\cong O(1).
\end{equation}
This action is a generalization of the action which is proposed in the Einstein gravity for some
dilaton fields \cite{dani}. We assume that the free
functions $f, U$ are analytic in $\Phi$ in the limit of large $\Phi$. Indeed this action is power-counting renormalizable\cite{power}. This power-counting renormalizability is valid for any polynomial form of the interaction coupling $f,U$. Comparing (\ref{g}),(\ref{g2}) we find there exists a gauge freedon for choice the field $\Phi=b(r)$. One trivial gauge is the constant gauge given by (\ref{b}). Thus our ansatz can be interpreted asa gauge free term in the effective action. Now we back to the analytic investigation of the solutions. Without loss of any generalization,we assume the isotropic ansatz for the matter distribution
\begin{equation}
p_t=p_r.\label{ansatz}
\end{equation}
The expression for  rotation curves of galaxies is \cite{v}
\begin{equation}
v_\varphi=\Big(\frac{r(e^a)'}{2e^a}\Big)^{1/2},\label{v}
\end{equation}
where prime denotes differentiation with respect to radial
coordinate $r$. This formula is the same as the Einstein gravity. We
must clarify this point here. The path of the free particle can be
obtained using the usual minimization method of the action for a
free particle
$$
\delta \int ds=0,\ \ ds=\sqrt{g_{\mu\nu}\dot x^{\mu}\dot
x^{\nu}}d\eta.
$$
Here $\eta$ is  the affine parameter. Using this equation we obtain
the following geodesic equation,
$$
\ddot x^{\mu}+\Gamma^{\mu}_{\nu\alpha}\dot x^\nu \dot x^\alpha=0.
$$
Here $\Gamma^{\mu}_{\nu\alpha}$ is the Levi-Civita connections,
defined by the symmetric part of the general connection from the
metricity equation. The asymmetric part of the connection
$\gamma^{\mu}_{\nu\alpha}$ has no portion in this geometrical
equation. Thus if we mean by the geodesic equation, the non auto
parallel motion, the same expression can be used. But using the auto
parallel formalism is another story and we will not enter in it. It
is easy to show that this geodesic equation is equivalent by the
Euler-Lagrange equations, derived from the following point like
Lagrangian for the test particle
$$
2\mathcal L=e^a \dot t^2-e^b\dot r^2-r^2
(\dot\theta^{2}+\sin^{2}\theta \dot\phi^{2}).
$$
The purely radial equation for test particle reads
\begin{eqnarray}
\dot r^2+U(r)=0,\ \
U(r)=e^{-b(r)}\Big(e^{-a(r)}\epsilon^2-\frac{h^2}{r^2}-1\Big),\ \
 h^2=p_{\theta}^2+\frac{p_{\varphi}^2}{\sin\theta^2}.
\end{eqnarray}
Here $p_{\theta},p_{\varphi}$ are the conjugate monenta of the
corresponding coordinates, $\epsilon=\dot t e^a$ is the energy
(local). Here we are assuming that there exist ZAMO (zero angular
momentum ) observer,  located at the spatial infinity
$r\rightarrow\infty$. We note that the tangential (rotation)
velocity reads
$$
v_{\varphi}=\sqrt{\frac{e^{a(r)} h^2}{r^2 \epsilon^2}}=\frac{r
a'(r)}{2},
$$
which coincides with (\ref{v}). Making use of isotropic ansatz
(\ref{ansatz}) in (\ref{presr}) and (\ref{prest}), we obtain
\begin{equation}
e^{a(r)}= \frac{r^{3-\sqrt{13-4e^c}}}{16(-13+4e^c)^2}(C_1 r^{\frac{\sqrt{13-4e^c}}{2}}
-C_2)^4,\label{ea}
\end{equation}
where $C_1$ and $C_2$ are new constants of integration.  A simple
direct calculation of torsion using (\ref{te}) and with metric
function given by (\ref{ea}) shows that in this case $T(r)\neq0$
(See figure 1).  By substituting (\ref{ea}) in (\ref{v}) we can plot
the rotation (tangential) velocity as it is plotted in figure 2. We
used from a large set of data come from the local tests of the
$f(T)$ models based on the cosmographic description
\cite{cosmography}. Figure 1 resembles the rotation curves for large
(spiral) galaxies \cite{rotv}. The scale where the velocity profile
$v_\varphi$ attains is $v_\varphi\approx 10^{-3}$, which roughly
corresponds to $r\approx300km/s$, in reasonable agreement with the
data. It is important to mention that the velocity profile shown in
figure 2, is constructed  basically on a phenomenological toy model.
It is gratifying that our $f(T)$ theory predicts a good velocity
profile that was argued to be a good phenomenological fit to the
data.
\begin{figure}[thbp]
\includegraphics[scale=0.5]{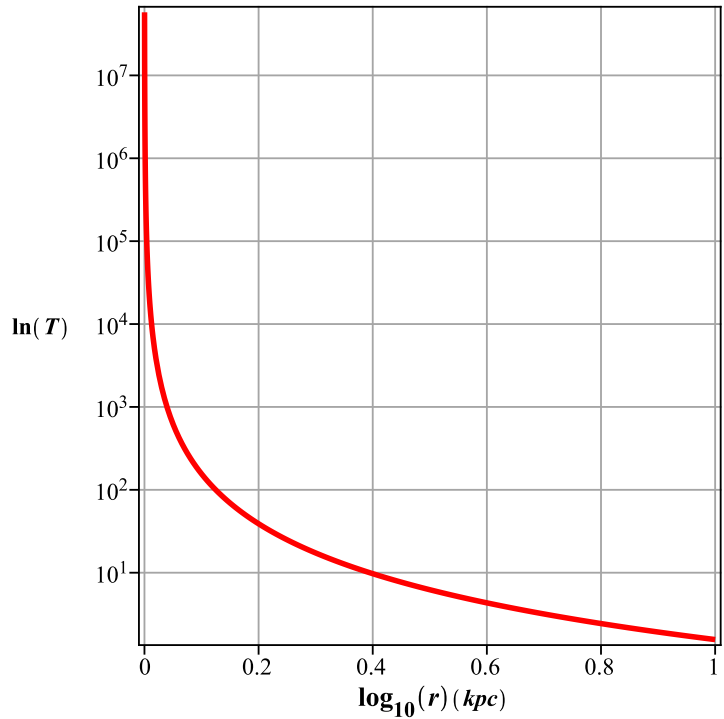}
\caption{  Non-zero scalar torsion $T$ for metric function given by
(\ref{ea}). }
\end{figure}

\begin{figure}[thbp]
\includegraphics[scale=0.5]{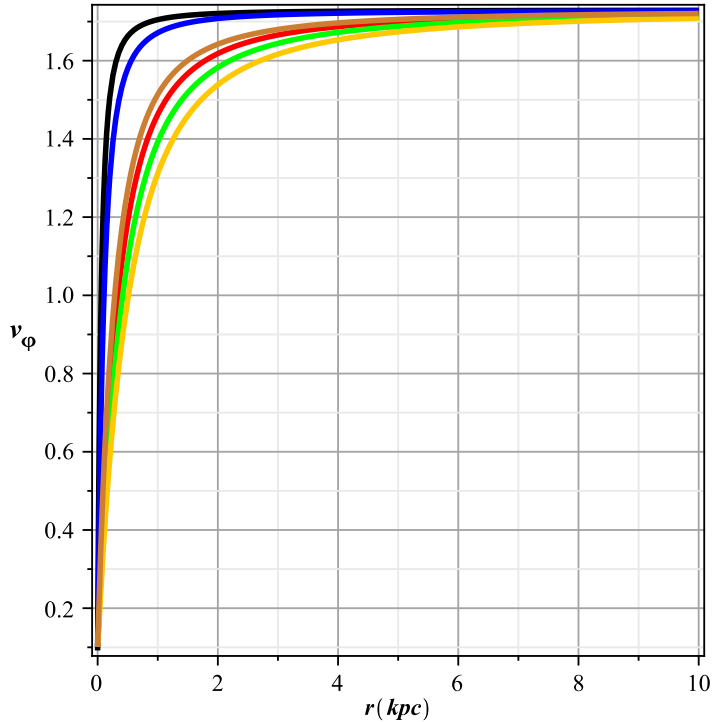}
\caption{  Toy model rotation curve for large spiral galaxy based on
our $f(T)$ model. The units of the vertical axis must be read by as factor  $\times 10 \frac{Km}{s}$. }
\end{figure}

\begin{figure}[thbp]
\includegraphics[scale=0.5]{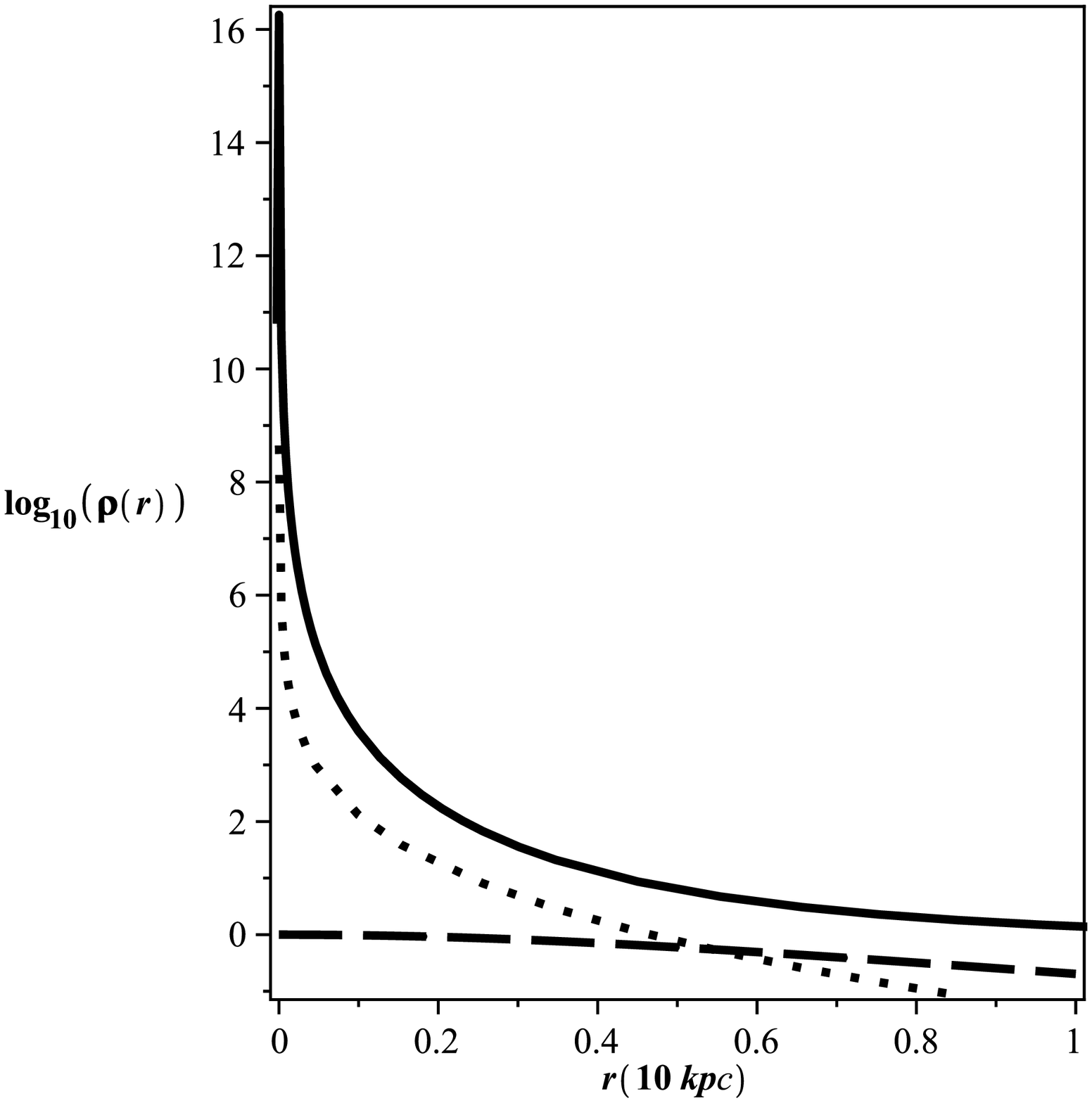}
\caption{  Two astrophysical halo density profiles in comparison to
our toy model. Solid is our prediction given by (\ref{ourmodel}),
dot is the density of NFW model (\ref{nfw}) and dash the estimation
of pseudo-isothermal sphere approximation introduced by
(\ref{model2}). The behavior of our model is very close to the
estimations of the NFW model \cite{rho}. }
\end{figure}

%%%%%%%%%%%%%%%%%55
\section{On central density profile $\rho(r)$ in $f(T)$ model}
%%%%%%%%%%%%%%%%%%%%%%
From observational  data we know that there exists a core with
roughly constant density (mass density) in the galaxy. Many models
have been proposed for this mass profile density. In brief these
models are used in the numerical simulations, specially for $\Lambda
$CDM:
\begin{eqnarray}
\rho(r)=\frac{\rho_i}{(\frac{r}{r_s})(1+\frac{r}{r_s})^2}\label{nfw}\\
\rho(r)=\frac{\rho_0}{1+(\frac{r}{r_c})^2}\label{model2}.
\end{eqnarray}
Here $\rho_i$ represents the density of Universe at the collapse
time, $ \rho_0$ is central density of the halo, $r_s$ is a
characteristic radius for the halo and $r_c$ is the radius of the
core \cite{rho}. Further, very close to the center, this density
profile is characterized by $r^n$ where $n\approx-1$, which means we
have a density cusp. The observations, often favor $n\approx0$, i.e.
a constant-density core.  Now we want to compare our estimation on
the form of the $\rho(r)$ from model of $f(T)$ by metric function
(\ref{ea}). Substituting (\ref{ea}) in (\ref{dens}) we obtain
\begin{eqnarray}
\rho(r)=\frac{1}{16\pi}\Big[c_{2}+\frac{2c_1(1-e^{-c})}{r^2}\Big]\label{ourmodel}
\end{eqnarray}
This is the exact mass profile density of our model of DM, which is
comparable with the two recently proposed models of the density
(\ref{nfw},\ref{model2}) (See the figure 3 for a comparison of the
estimated model of us given in (\ref{ourmodel}) and two models
(\ref{nfw},\ref{model2}) from  astrophysical data given from
\cite{rho}).

%%%%%%%%%%%%%%%%%%%%%
\section{Conclusion}
%%%%%%%%%%%%%%%%%%%%%
In this letter we obtained the rotation curve of the galaxies  in
the $f(T)$ gravity. We proposed that the galaxy metric remains
spherically symmetric and static. Then by solving the general
equations of the metric components we obtained the lapse function
$N=e^a$ as a function of the radial coordinate $r$. Our qualifying
discussions shown a very good agreement between the rotation curves
in this model and other curves obtained from the data. The scale
where the velocity profile $v_\varphi$ attens is $v_\varphi\approx
10^{-3}$, which roughly corresponds to $r\approx300$ km/s, in
reasonable agreement with the data. Further, the exact mass profile
density of our model of DM,  is in good agreement with the two
models of the density. It proves that dark matter problem can be
resolved as the effect of the torsion of the space time easily.

\textbf{It is well-known from ECSK theory that torsion couples to
the spin of matter, hence one can infer to measure the torsion
produced by any massive spinning object (including massive spiral
and elliptical galaxies). However there has been assumptions about
testing these theories, such as ``all torsion gravity theories
predict observationally negligible torsion in the solar system,
since torsion (if it exists) couples only to the intrinsic spin of
elementary particles, not to rotational angular momentum''
\cite{tegmark}. Mao \textit{et. al.} have shown that Gravity Probe B
is an ideal experiment for constraining several torsion based
theories \cite{tegmark}. Although their analysis is based on the
torsion field around a uniformly rotating spherical mass such as
Earth, the task of constraining torsion around massive galaxies is
still open for exploration.}


\begin{thebibliography}{99}
\bibitem{power}
  This calculation is done most conveniently in the gauge
theoretic formulation analog to \cite{visser}. We will present the details in the forthcoming paper on renormalizability of the $f(T)$ models.
\end{thebibliography}
\end{document}